\documentstyle[epsf,rotate,12pt]{ioplppt}
\begin{document}
\title{Suppression of level hybridization due to Coulomb interactions}
\author{Pascal Cedraschi and Markus B\"uttiker}
\address{D\'epartement de Physique Th\'eorique,
Universit\'e de Gen\`eve,\\
24, quai Ernest Ansermet, CH-1211 Geneva 4, Switzerland}
\date{\today}
\begin{abstract}
We investigate an ensemble of systems formed by a ring enclosing a
magnetic flux. The ring is coupled to a side stub via a tunneling 
junction and via Coulomb interaction. We generalize the notion
of level hybridization due to the hopping, which is naturally
defined only for one-particle problems, to the many-particle case,
and we discuss the competition between the level
hybridization and the Coulomb interaction.
It is shown that strong enough Coulomb interactions
can isolate the ring from the stub, thereby increasing the persistent
current. Our model describes a strictly canonical system (the
number of carriers is the same for all ensemble members).
Nevertheless for small Coulomb interactions and a long side stub
the model exhibits a persistent current typically 
associated with a grand canonical ensemble of rings 
and only if the Coulomb interactions are sufficiently strong 
does the model exhibit a persistent current which one expects 
from a canonical ensemble. 
\end{abstract}
\maketitle

\section{Introduction}
Hybridization of levels 
or elementary excitations is encountered in many problems of
physics \cite{heitler:bonding,stueck,fulde,vaart,
waugh:multidot,pfannkuche:artmol}
and chemistry
\cite{larsson:eltransfer,marcus:eltransfer,skourtis:eltransfer}
in which two weakly coupled subsystems interact with one another. 
In this work we are interested in the hybridization of electronic levels
and in particular in the case in which a passage through the 
hybridization point is associated with the displacement of charge in real
space. If hybridization invokes the transfer of charge between two 
weakly coupled systems the Coulomb interaction can be expected 
to play a dominant role: The charge transfer 
is permitted only if it is associated 
with a charge distribution which 
exhibits a smaller interaction energy 
than the original configuration. If this is not the case 
the Coulomb interaction can be expected to 
suppress the hybridization. 

\begin{figure}
\centerline{\rotate[l]{\text{\epsfysize=10cm\epsfbox{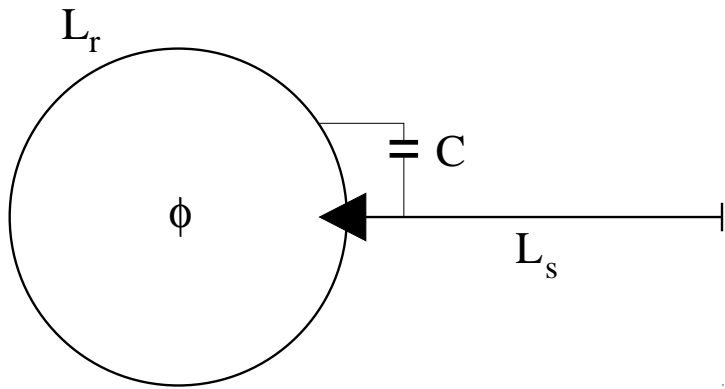}}}}
\caption{\label{sys}A ring pierced by a flux $\phi$ and
coupled to a (finite length) side stub. The triangle represents the
a three way tunneling junction and the Coulomb interaction is taken into account
via the capacity $C$.}
\end{figure}

To investigate this question we consider a simple model system 
shown in figure~\ref{sys}. A ring pierced by an Aharonov-Bohm 
flux \cite{bil} is weakly 
coupled to a side-branch \cite{buettiker:ringstub}.
Of interest 
are the charge transfer processes between the ring and the stub 
and their effect on the persistent current. The simple model 
shown in figure~\ref{sys} permits to investigate the interplay 
between highly mobile electron states and states in which 
the electron is localized. In this model, a state in which the 
electron is predominantly in the ring is very sensitive to the 
flux and provides a strong contribution to the persistent current 
whereas a state in which the electron is predominantly localized in 
the stub is nearly insensitive to a variation of the flux.
In the absence of Coulomb interaction the hybridization of these
two type of states leads to a small persistent current.
If now the Coulomb interactions are switched on, transfer of charge 
into and out of the localized states is generally not energetically
favourable. As a consequence the model exhibits an increased 
persistent current in the presence of Coulomb interaction. 
For a sufficiently strong Coulomb interaction the presence of the 
side branch is irrelevant, the persistent current is that of a loop
without a side branch. Thus
the larger persistent current can be viewed as a consequence 
of the suppression of hybridization of highly mobile states with 
localized states.
It is well appreciated 
that many-ring experiments \cite{levy:manyrings}
as well as experiments on single metallic diffusive
rings \cite{chandrasekhar:singlering}
yield a value for the persistent
current which is much larger than predicted by theories
which neglect interactions. In contrast the measurements on
single ballistic semiconductor rings \cite{mailly:singlering}
seem in accord even with the 
predictions of non-interacting theories. 
Possibly discussions of the persistent current in the absence
of interactions find a very small value for the current 
since hybridization of levels is not inhibited. 
In the absence of interactions arbitrary charge distribution patterns 
are permitted. 
If, as is shown in this work, Coulomb interactions 
can effectively inhibit level hybridization
of flux sensitive states with flux-insensitive localized states, this might 
offer a mechanism which permits the much larger currents 
observed in experiments and in discussions which take 
Coulomb interactions into account \cite{ambegaokar}. 
The discussion presented here
is limited to the simple example shown in figure~\ref{sys}
and does not address the case of metallic diffusive systems.
We point out, however, 
that the work of Pascaud and Montambaux \cite{montambaux:network} 
which considers metallic diffusive conductors with a geometry
similar to the one of interest here, gives results which are in
accord with the findings reported below. 

The model system, proposed in \cite{buettiker:ringstub},
has been analyzed in a number of works. Reference
\cite{buettiker:ringstub} treated interactions on the Hartree
level in a random phase approximation. A discussion which takes 
the effect of charge quantization 
in the charge transfer explicitly into account 
and treats interactions within a quantum coherent 
charging model was subsequently provided
by Stafford \cite{buettiker:ringdot:prl,buettiker:ringdot} in collaboration
with one of the authors. A computational investigation in which interactions
are treated in a Hubbard like model is provided by Anda
et al.~\cite{anda:bistab}. Without interactions the geometry
of figure~\ref{sys}
represents an elementary system which has also found
some interest \cite{deo}.  
There are a number of closely related geometries: 
First instead of a ring and a side branch one can
investigate two coupled rings 
as has been done by Canali et al.~\cite{canali:doublering}.
Alternatively, two quantum dots in parallel \cite{akera:2dots}
or in series \cite{pfannkuche:artmol,pals:coherenttunneling,
matveev:doubledot,gold,izumida:2dots}
usually attached to leads to investigate transport,
have been considered. 
A system very similar to ours, with the side stub replaced by
a quantum dot is investigated
in \cite{buettiker:ringdot:prl,buettiker:ringdot}. These authors also
treat the case where the quantum dot is incorporated into one of
the arms of the ring and where dot and ring 
are each separately coupled to a gate.
Throughout we focus on 
the closed system of figure~\ref{sys} in which the level 
spectrum is discrete. If the stub attached to the ring is not closed
but connected to a reservoir \cite{buettiker:wavesplitter} 
we have an open grand canonical system 
in which the spectrum of the wire connecting the reservoir and the ring is 
continuous. 
Effects of Coulomb charging in such models have been discussed by 
Beenakker and van Houten \cite{beenakker} and by
Moskalets \cite{moskalets}.     
Charge transfer between a Luttinger liquid ring 
and a reservoir is discussed by Sandstr\"om
and Krive \cite{sandstroem:luttingerring}
and Krive et al.~\cite{krive:luttingertd}. 
 
The purpose of the present work is to extend the discussion of 
\cite{buettiker:ringdot:prl,buettiker:ringdot} which focused on 
quantum coherent resonant charge transfer between subsystems to 
the case of off-resonant charge transfer. 
The main difficulty in further advancing the notions put forth 
above is that level hybridization is a single particle 
concept. Thus we can only compare the different properties of the interacting 
system with that of the non-interacting system
and show that the signatures of level hybridization
encountered in the non-interacting system 
disappear with increasing interaction strength. 
To demonstrate this it is necessary to investigate charge 
transfer not only under resonant conditions 
but in a wide parameter range of the model. 
Furthermore it is necessary to consider an ensemble of
systems rather than a single system. 
It is well known that the size of the persistent currents 
depends sensitively on the ensemble 
considered \cite{bouchiat:canvsgrand,kamenev:ensemble}.
The present model, although it is strictly a canonical one, since 
the overall charge is conserved, nevertheless, as far as the persistent 
current is concerned, shows aspects 
usually associated with a grand canonical ensemble in the limit of
weak interactions and for a sufficiently long stub.

\section{\label{sec:model}The model}
We consider the following model.
A ring is coupled to a side stub via particle hopping and
via electrostatic energy.
In the absence of coupling between the ring and the stub 
both subsystems are perfect, disorder free conductors. 
The energy levels of the ring are denoted by
$\epsilon_n^{(r)}(\phi)$. They are periodic in the flux
$\phi$ with period $\phi_{0}= hc/e$. 
The eigenstates of the stub 
are extended along its entire length and their energies are 
denoted by  $\epsilon_n^{(s)}$.
For simplicity, the electrons are considered as spinless.
We introduce the operators 
$\hat{a}_i^\dagger $ 
which create an electron in state $i$ of the ring 
and the operators $\hat{b}_j^\dagger$ which create an
electron in the stub.
The Hamiltonian is the sum of the kinetic energy $\hat{K}$ of the electrons 
in the ring and the stub, the hopping energy $\hat{\Gamma}$,  
and the Coulomb energy $\hat{H}_C$. With the energies and operators
introduced above we have a kinetic energy
\begin{equation}
\label{kin}
\hat{K}=\sum_i\epsilon_i^{(r)}(\phi)\hat{a}_i^\dagger\hat{a}_i
+\sum_j\epsilon_j^{(s)}\hat{b}_j^\dagger\hat{b}_j
\end{equation}
and a coupling energy
\begin{equation}
\label{gamma}
\hat{\Gamma}=\sum_{i,j}\left( t_{ij}\hat{a}_i^\dagger\hat{b}_j+ h.c.\right).
\end{equation}
The one-particle spectra $\epsilon_i^{(r)}$ and $\epsilon_j^{(s)}$
of the ring and the stub are given by the spectra of the free
particle. If the particle numbers $N_r$ in the ring and $N_s$
in the stub are large, the spectra 
may be linearized. Denoting the velocity of the topmost occupied state
in the stub by $v_F^{(s)}$, we find for a stub of length $L_s$ 
a level spacing 
\begin{equation}
\label{delta} 
\Delta ={\pi\hbar v_F^{(s)}\over L_s}. 
\end{equation}
We characterize the spectrum of the 
ring of circumference $L_r$ by the width $w$ of the levels 
which they obtain as a function of flux.
For a ballistic ring in which the topmost state has a a velocity 
$v_F^{(r)}$ the level width is given by 
\begin{equation}
\label{w}
w\equiv|\epsilon_{N_r}^{(r)}(\phi_0/2)
-\epsilon_{N_r}^{(r)}(0)|={\pi\hbar v_F^{(r)}\over L_r}. 
\end{equation}
Below we frequently use 
$\Delta$ and $w$ to characterize the spectrum of the system. 

The Coulomb energy is taken into account with the help 
of a geometrical capacitance $C$ and is obtained as follows: 
The $N_{r}$ electrons on the ring,
$N_{s}$ electrons on the stub are held in place 
by an ionic background charge $eN_{r}^{+}$ on the ring 
and $eN_{s}^{+}$ on the stub. 
We consider a system that is overall charge neutral 
and therefore, $N = N_{r}+N_{s} = N_{s}^{+}+N_{r}^{+}$.
The Coulomb energy of these charges is 
$E_C = (1/2) [(N_{r} -N_{r}^{+}) eU_{r} + (N_{s} -N_{s}^{+})eU_{s}].$
Relating the charge imbalance $Q = e (N_{r} -N_{r}^{+})$
to the potential difference $U_r-U_s $ via the geometrical capacitance $C$,
$Q =C(U_r-U_s)$, and using the charge neutrality condition 
gives
\begin{equation}
\label{E_c}
\hat{H}_C=\frac{e^2}{2C}\left(\hat{N}_r-N_r^+\right)^2\, ,\quad
\hat{N}_r=\sum_i \hat{a}_i^\dagger\hat{a}_i\, .
\end{equation}
Note that this energy is equal to
$\hat{H}_C=\frac{e^2}{2C}(\hat{N}_s-N_s^+)^2$.
Below we discuss various simple limits of this Hamiltonian. 

\begin{figure}
\centerline{\text{\epsfysize=10cm\epsfbox{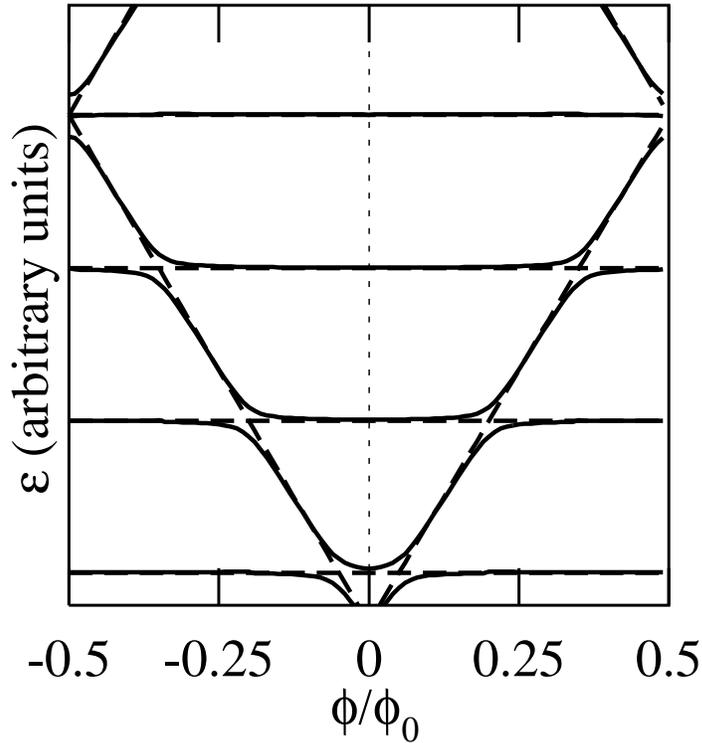}}}
\caption{\label{spectrum}Part of the single particle energy spectrum 
of a ring connected to a stub (full lines).
The dashed lines represent the spectra of the uncoupled ring (levels
21 and 22) and stub (levels 30 to 33), respectively.
The spectrum of the Hamiltonian including tunneling was calculated by
matrix diagonalization with
a level width $w$ in the ring, a level spacing $\Delta=2w/3$ in the stub
and a coupling energy $|t|=0.1 w$.}
\end{figure}

For vanishing Coulomb interaction ($e^2/2C=0$) the problem reduces
to a one-particle problem which we discuss now.
In the absence of hopping between the ring and the stub ($\hat{\Gamma} = 0$)
the spectrum consists of the flux sensitive spectrum of a 
perfect ring and a completely flux insensitive spectrum 
of the stub. This spectrum is indicated 
by broken lines in figure~\ref{spectrum}. 
From their flux dependence the ring states behave like 
extended states, whereas the stub states behave like completely 
localized states. 
If we now turn on the coupling between 
the stub and the ring the states of these two subsystems 
hybridize. Instead of an intersection of two nearby levels 
a gap of the order of $|t|$ opens up. 
The solid lines in figure~\ref{spectrum} represent  
the spectrum of the ring and the stub 
for a finite coupling strength $|t|$ obtained by matrix diagonalization.
As a consequence the wave function describing 
a carrier initially in an (extended) ring state changes its character 
at the hybridization point and turns
into a wave function which describes a carrier localized on the stub. 
As we pass the hybridization point a carrier is thus transferred from the 
ring into the stub. Separated in energy by a gap of the
order of the coupling energy 
$|t|$ there is a second wave function 
which describes the transfer of a carrier initially 
localized on the stub into the ring as we pass the 
the hybridization point. For weak coupling and if both 
states are occupied the net charge transferred vanishes. 
Thus for weak coupling the effect of hybridization is only 
relevant for the topmost occupied level. 

It is the purpose of this work to investigate 
the suppression of the hybridization 
when the Coulomb energy is present. 
Clearly, if the transfer of the topmost electron from the ring to the 
stub leads to a marked deviation of charge distribution away from 
a locally electroneutral solution the charge transfer will effectively 
be inhibited in the presence of strong Coulomb interaction. 
The complication which is encountered in the characterization 
of this phenomenon arises from the fact that level 
hybridization, as it is typically discussed and shown in
figure~\ref{spectrum},
applies to a non-interacting system. In an interacting system we can 
in general not follow single particle levels. Thus we can discuss the 
suppression of hybridization only indirectly by showing 
that the features in the ground state energy which are typical
for hybridization in the large capacitance limit 
vanish with increasing Coulomb interaction. 
Alternatively, we can investigate properties of the ground state
wave function,
in particular the fluctuations of the charge on the ring
as a function of the Coulomb interaction. 

Therefore we discuss now the behaviour of our system in the case
where the charging energy is non-vanishing. In a first step,
the tunneling energy is neglected.
This is the limit of the standard incoherent Coulomb 
blockade description \cite{averin:qdot,beenakker:CB}.
It permits us to discuss the 
charging states of the model. Later on we will include the 
quantum mechanical transmission to find small corrections
to this quasi-classical picture.  
At zero temperature
the free energy is given by 
\begin{equation}
\label{free_en}
F(N_r,\phi)=\sum_{n=1}^{N_r}\epsilon_n^{(r)}(\phi)
+\sum_{m=1}^{N_s}\epsilon_m^{(s)}
+\frac{e^2}{2C}\left( N_r-N_r^+\right)^2\, .
\end{equation}
The operator $\hat{N}_r$ has been replaced by its eigenvalue $N_r$, and
$N_s=N-N_r$. 
For a given flux $\phi$, the state with 
$N_r$ particles on the ring is realized 
if its free energy is smaller than all the other free energies
obtained with different particle numbers $N_r^\prime$. 
It is clear that for very large Coulomb energies the state with
$N_r =N_{r}^{+}$ for which the electronic charge 
exactly compensates the ionic background charge will be the 
state with minimal free energy. The Coulomb energy of this state
vanishes, whereas for the neighbouring states $N_{r}^{+} \pm 1$
we have to pay an additional energy $e^2/2C$. 
Thus for large Coulomb energies we have 
$F(N_r^+,\phi)\ll F(N_r^+\pm 1,\phi)$.
If we now lower the Coulomb energy either 
$F(N_r^++1,\phi)$ or $F(N_r^+-1,\phi)$ might become 
smaller than $F(N_r^+,\phi)$. 
Thus a charge transfer into or out of the ring occurs
at the points of energetic degeneracies for which 
\begin{equation}
\label{hybridization}
F(N_r^+,\phi)=F(N_r^+\pm 1,\phi).
\end{equation}
Using (\ref{free_en}) yields
\begin{eqnarray}
\label{hybridization1}
\epsilon_{N_r^++1}^{(r)}(\phi)-\epsilon_{N_s^+}^{(s)}+\frac{e^2}{2C}=0
&\quad&\text{for ``+''}\\
\label{hybridization2}
\epsilon_{N_s^++1}^{(s)}-\epsilon_{N_r^+}^{(r)}(\phi)+\frac{e^2}{2C}=0
&\quad&\text{for ``$-$''}
\end{eqnarray}
with $N_r^++N_s^+=N$.
Since the energies of the ring depend on the flux these equations 
might have a solution only for a particular flux $\pm \phi^{*}$.
At this flux the topmost filled stub state has an
energy $\epsilon_{N_s^+}^{(s)}$
which is equal to the first empty state of the ring 
$\epsilon_{N_r^++1}^{(r)}(\phi^{*})$ augmented by the Coulomb energy
$e^2/2C$. Alternatively the topmost occupied ring state
$\epsilon_{N_r^+}^{(r)}(\phi^{*})$ has an energy which is 
equal to the topmost empty stub state augmented by the charging energy. 
In the limit of infinite capacitance the flux $\phi^{*}$
corresponds to a hybridization point in figure~\ref{spectrum}.
In the quasi-classical model which neglects phase coherence between
ring and stub states the hybridization region has zero extend:
The charge on the ring jumps as the flux is moved through $\phi^{*}$.
Quantum mechanically, if phase coherence is taken into account 
hybridization extends over a range of flux which is determined both 
by the the strength of the tunneling matrix element $|t|$
and the magnitude of the Coulomb energy $e^2/2C$.
In the next section we will describe these transitions
in a quantum coherent model
in more detail. 

By biasing this system via
gates \cite{buettiker:ringdot:prl,buettiker:ringdot}, 
one can essentially create an arbitrary
charge imbalance, i.e.\ $N_r\neq N_r^+$. 
In this case the Hamiltonian
(\ref{kin}--\ref{E_c}) contains additional terms which describe the coupling 
of the system to the gate voltage. In such a case the system can exhibit 
resonances like in (\ref{hybridization})
at any strength $e^2/2C$
of the interaction depending only on the 
value of the gate voltage.  For a ring connected via a wire 
to a reservoir, capacitance fluctuations have been
discussed by Gopar, Mello and B\"uttiker \cite{gopar} and by  
Aleiner and Glazman \cite{aleiner:CQ}.
For large charging energy this Hamiltonian can
be cast into a Kondo-like form \cite{kondo}. The Kondo-like features
of this problem have been treated by
Matveev \cite{matveev:kondo}.
As the system we are discussing does not contain gates, the resonance
condition (\ref{hybridization}) cannot be fulfilled for large
charging energies $e^2/2C$. Therefore it does not exhibit a
Kondo-effect.

\section{Quantum corrections}

In this section we the treat quantum corrections to the classical 
picture developed above. Different treatments apply depending on whether
or not the charge transfer is resonant or off-resonant. 

\subsection{\label{sec:resonant}Resonant charge transfer}

\begin{figure}
\centerline{\text{\epsfysize=10cm\epsfbox{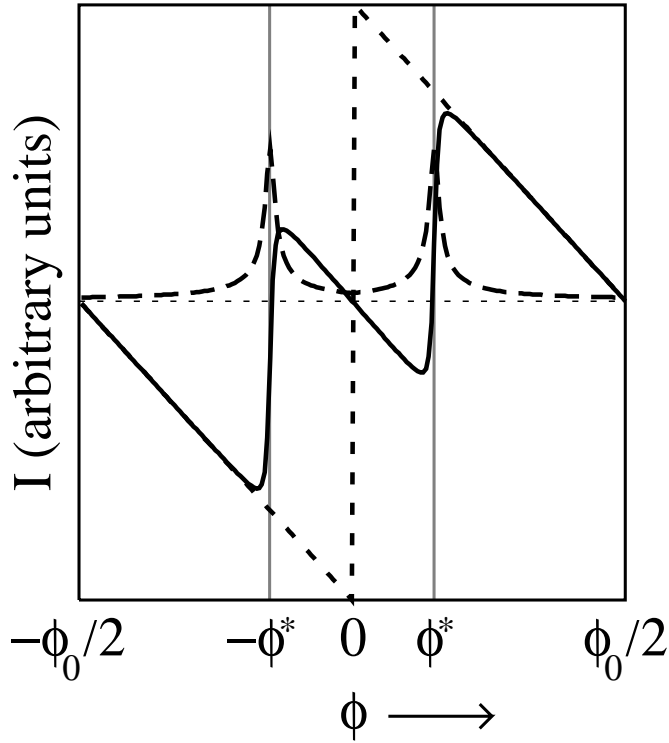}}}
\caption{\label{fig:pc}The persistent current (solid line) and
its fluctuations $(\Delta I^2)^{1/2}$ (dashed line) in the
two-level approximation with $N_r^+$ even and $|t|\approx 0.32w$.
For comparison 
the persistent current for $N_r^+$ electrons in the ring and
in the absence of hybridization is shown (short dashed line).
The persistent current vanishes near the
hybridization points $\pm\phi^*$ ($=\pm\phi_0/6$ in the example)
whereas the fluctuations have maxima near the same points,
which are of the same order
of magnitude as the persistent current in the absence of
hybridization.}
\end{figure}

Hybridization occurs when one of the equations (\ref{hybridization1},
\ref{hybridization2}) is fulfilled for some $\pm\phi^*$.
Without loss of generality we may assume that it is
equation (\ref{hybridization2}) that is fulfilled.
As shown in recent works \cite{buettiker:ringdot:prl,buettiker:ringdot},
in the weak tunneling limit
it suffices to consider hybridization between a
state in the ring and a state in the stub.
Then the Hamiltonian
reduces to a $2\times 2$ matrix
\begin{equation}
\label{H_h}
\hat{H}_h=\left(\matrix{
\epsilon_{N_r^+}^{(r)}(\phi) & t\cr
t^* & \epsilon_{N_s^+ +1}^{(s)}+\frac{e^2}{2C}}\right)
+\left(\sum_{n=1}^{N_r^+ -1}\epsilon_n^{(r)}(\phi)
+\sum_{m=1}^{N_s^+}\epsilon_m^{(s)}\right){\bf 1}
\end{equation}
that is easily diagonalized.
The eigenvalues of $\hat{H}_h$ show a gap around the hybridization point
$\phi=\pm\phi^*$. It is a typical hybridization effect, and gaps of the
same kind also open up in the one-particle spectrum of the 
Hamiltonian $\hat{K}+\hat{\Gamma}$ (equations~\ref{kin},~\ref{gamma}),
see figure~\ref{spectrum}.
The eigenstates of $\hat{H}_h$ are not eigenstates of the
(reduced) particle number operators $\hat{N}_r$ (and $\hat{N}_s$):
$\hat{N}_r=\hat{N}_h+(N_r^+ -1){\bf 1}$ with
\begin{equation}
\hat{N}_h=\left(\matrix{ 1 & 0\cr 0 & 0 }\right).
\end{equation}
Around $\phi=\pm\phi^*$ the topmost electron is extended over both
subsystems. As a consequence and we find in the ground state an 
expectation value of the charge $e\langle\hat{N}_r\rangle$ which is
not an integral multiple of the elementary charge. Furthermore
there are strong particle number fluctuations $\Delta N_r{}^2$
away from the average $\langle\hat{N}_r\rangle$.  
The particle number fluctuations \cite{buettiker:ringdot}
are determined by 
\begin{equation}
\label{numberfluct}
\Delta N_r{}^2=\left\langle(\hat{N}_r-\langle\hat{N}_r\rangle)^2\right\rangle
={|t|^2\over\left(\epsilon_{N_s^++1}^{(s)}-\epsilon_{N_r^+}^{(r)}(\phi)
+{e^2\over 2C}\right)^2+4|t|^2}\, .
\end{equation}
Note that at the hybridization point $\epsilon_{N_s^+ +1}^{(s)}
-\epsilon_{N_r^+}^{(r)}(\phi^*)+e^2/2C=0$ we have the maximal fluctuation
$(\Delta N_r{}^2)^{1/2}=1/2$.
The persistent current in the ring is given by $I(\phi)=-c\,\partial F/
\partial\phi$, where $F=\langle\hat{H}\rangle$. In the
two-level approximation, it reads
\begin{equation}
\label{eq:pc}
I(\phi)=-c{\partial\epsilon_{N_r^+}^{(r)}\over\partial\phi}
\langle\hat{N}_h\rangle +I_{N_r^+-1}(\phi).
\end{equation}
The total persistent current consists of a contribution 
of the topmost ring level with an average 
occupation $\langle\hat{N}_h\rangle$ and
of the full persistent 
current $I_{N_r^+-1}(\phi)$ of all $N_r^+ -1$ ring states 
below the the topmost level. This form of the persistent current is a
consequence of the weak coupling between the ring and the stub. 
In this case all ring levels below the topmost level are fully occupied 
and give rise to a persistent
current $-c\sum_{n=1}^{N_r^+-1}\partial\epsilon_n^{(r)}/\partial\phi
=I_{N_r^+-1}(\phi)$. 
$I(\phi)$ vanishes near the hybridization points $\phi=\pm\phi^*$,
as discussed in \cite{buettiker:ringstub}.
It is interesting to observe what happens to the quantum mechanical
current fluctuations when the persistent current itself goes to zero.
They are given by~\cite{buettiker:ringdot}
\begin{equation}
\label{eq:pcfluct}
\langle\Delta I^2\rangle=c^2\left(
{\partial\epsilon_{N_r^+}^{(r)}\over\partial\phi}\right)^2
\langle\Delta N_r{}^2\rangle.
\end{equation}
They are maximal near $\phi=\pm\phi^*$ where
$I(\phi)$ vanishes, and there
$\langle\Delta I^2\rangle^{1/2}$ is of the same magnitude
as $I(\phi)$ is away from $\phi=\pm\phi^*$.
The behaviour of $I(\phi)$ and $\Delta I^2$ is summarized
in figure~\ref{fig:pc}.

An interesting quantity that is sensitive to hybridization is the flux induced 
capacitance \cite{buettiker:ringstub,buettiker:ringdot:prl,buettiker:ringdot}
$C_\phi=e\,\partial\langle\hat{N}_r\rangle /\partial\phi$.
This quantity determines the charge increment (in the ring) 
in response to a small variation in the flux much like 
an electrochemical capacitance coefficient determines 
the increment in charge in response to a variation of a gate voltage. 
The flux induced capacitance is a particularly interesting quantity 
to consider here since it exhibits a large 
resonant-like spike at a hybridization point $\phi=\pm\phi^*$. 
In the two-level approximation considered here the 
resonance is determined \cite{buettiker:ringdot:prl,buettiker:ringdot} by 
\begin{equation}
C_\phi=
-{4|t|^2\partial\epsilon_{N_r^+}^{(r)}/\partial\phi\over
\left[\left(\epsilon_{N_s^+ +1}^{(s)}-\epsilon_{N_r^+}^{(r)}(\phi)+
{e^2\over 2C}\right)^2+4|t|^2\right]^{3/2}}\, .
\end{equation}
At resonance $C_\phi = (\partial\epsilon_{N_r^+}^{(r)}/\partial\phi)/2|t|$. 
We will not investigate this quantity here any further but note 
that $cC_\phi$ has the dimension of conductance 
and in topological discussions of the Hall effect it is sometimes
taken to be the Hall conductance \cite{laughlin:hallcond}.  

If we are not close to a hybridization point, however, the two-level
approximation is insufficient to discuss the quantities of 
interest here. 

\subsection{\label{subsec:offres}Off-resonant charge transfer}

To discuss the transfer of charge away from the resonant points 
we decompose the Hamiltonian
(\ref{kin}--\ref{E_c}) as $\hat{H}=\hat{H}_0+\hat{\Gamma}$
and treat $\hat{\Gamma}$ as a perturbation. Note that
$\hat{H}_0=\hat{K}+\hat{H}_C$ contains the Coulomb energy term.
We assume in the following that there are no energetic degeneracies
(cf.\ equation~\ref{hybridization}).
A complete orthonormal system of eigenvectors to $H_0$ is constructed
as follows. Let $|\psi_0\rangle$ denote the vacuum (no particles) and
$I_m$, $J_n$ index sets with $m$, $n$ elements,
respectively. The basis reads
\begin{equation}
\label{ons}
|I_mJ_n\rangle=\prod_{i\in I_m}\hat{a}_i^\dagger
\prod_{j\in J_n} \hat{b}_j^\dagger
|\psi_0\rangle\, ,\quad m+n=N.
\end{equation}
We need only a small subset of them,
one of which is the ground state 
of the uncoupled system
\begin{equation}
|0\rangle=\prod_{i=1}^{N_r^+}\hat{a}_i^\dagger
\prod_{j=1}^{N_s^+}\hat{b}_j^\dagger|\psi_0\rangle
\end{equation}
with energy
\begin{equation}
E_0=\sum_{i=1}^{N_r^+}\epsilon_i^{(r)}(\phi)
+\sum_{j=1}^{N_s^+}\epsilon_j^{(s)}.
\end{equation}
We will calculate the corrections to the ground state energy
and the particle number fluctuations to second order perturbation
theory in $\hat{\Gamma}$ \cite{LanLif:3} (lowest non-vanishing order). 
We discuss briefly the states that contribute to the
ground state of the coupled system.
To first order in $\hat{\Gamma}$, one electron can hop from the
stub to the ring or vice-versa. We denote the states created from
$|0\rangle$ by one electron hopping by $|\alpha\rangle$. They
are of the form (\ref{ons}), but they are most easily expressed
in terms of the ground state of the uncoupled system
\begin{equation}
|\alpha\rangle=\left\{\matrix{
\hat{a}_i^\dagger\hat{b}_j |0\rangle\, ,\hfill
& \text{with } i>N_r^+ ,\; j\leq N_s^+\hfill\cr
\hat{b}_j^\dagger\hat{a}_i |0\rangle\, ,\hfill
& \text{with } i\leq N_r^+ ,\; j>N_s^+\hfill
}\right.\, .
\end{equation}
The corresponding eigenvalues are denoted by $E_\alpha$,
i.e.\ $\hat{H}_0|\alpha\rangle=E_\alpha|\alpha\rangle$.
To second order in $\hat{\Gamma}$ one has to consider processes that
involve two electrons, namely
\begin{enumerate}
\item one electron hopping from the ring to the stub and back, thereby
creating an excitation in the ring,
\item one electron hopping from the stub to the ring and back, thereby
creating an excitation in the stub,
\item one electron hopping from the ring to the stub, the other one
hopping from the stub to the ring, creating excitations in the ring
and in the stub,
\item two electrons hopping from the stub to the ring and
\item two electrons hopping from the ring to the stub.
\end{enumerate}
We use the summary notation $|\beta\rangle$ for
the states emerging from the ground state via these
processes. Formally (and in the same order as above)
\begin{equation}
|\beta\rangle=\left\{\matrix{
\hat{a}_k^\dagger\hat{a}_l |0\rangle ,\hfill
& \text{with } k>N_r^+,\; l\leq N_r^+ \hfill\cr
\hat{b}_k^\dagger\hat{b}_l |0\rangle ,\hfill
& \text{with } k>N_s^+,\; l\leq N_s^+ \hfill\cr
\hat{a}_k^\dagger\hat{a}_l\hat{b}_m^\dagger\hat{b}_n |0\rangle ,\hfill
& \text{with } k>N_r^+,\; l\leq N_r^+,\; m>N_s^+,\; n\leq N_s^+ \hfill\cr
\hat{a}_k^\dagger\hat{a}_l^\dagger\hat{b}_m\hat{b}_n |0\rangle ,\hfill
& \text{with } k>l>N_r^+,\; m<n\leq N_s^+ \hfill\cr
\hat{b}_m^\dagger\hat{b}_n^\dagger\hat{a}_k\hat{a}_l |0\rangle ,\hfill
& \text{with } k<l\leq N_r^+,\; m>n>N_s^+ \hfill}\right. ,
\end{equation}
and for the eigenvalue we write $\hat{H}_0|\beta\rangle
\equiv E_\beta|\beta\rangle$.

To second order in $\hat{\Gamma}$, the ground state energy reads
\begin{equation}
\label{gse}
E=E_0+\sum_\alpha {|\langle\alpha|\hat{\Gamma}|0\rangle |^2
\over E_0-E_\alpha},
\end{equation}
and the ground state $|\Omega\rangle$ of the coupled system
\begin{equation}
\label{gs}
\fl\left|\Omega\right\rangle=\left( 1- {1\over 2}\sum_\alpha
{|\langle\alpha |\hat{\Gamma}|0\rangle |^2\over (E_0-E_\alpha)^2}\right)
|0\rangle
+\sum_\alpha
{\langle\alpha |\hat{\Gamma}|0\rangle \over E_0-E_\alpha}
|\alpha\rangle
+\sum_{\alpha ,\,\beta}
{\langle\beta|\hat{\Gamma}|\alpha\rangle\langle\alpha |\hat{\Gamma}|0\rangle
\over (E_0-E_\alpha)(E_0-E_\beta)}|\beta\rangle.
\end{equation}
Summation over $|\alpha\rangle$ and $|\beta\rangle$ 
in (\ref{gse}, \ref{gs}) does
not include the ground state $|0\rangle$ of the uncoupled system.
Now we can also calculate the particle number fluctuations
\begin{equation}
\label{dN}
\Delta N_r{}^2=\langle\Omega|\hat{N}_r{}^2|\Omega\rangle
-\langle\Omega|\hat{N}_r|\Omega\rangle^2
=\sum_\alpha {|\langle\alpha|\hat{\Gamma}|0\rangle |^2
\over (E_0-E_\alpha)^2}\, .
\end{equation}
The states $|\beta\rangle$ do not appear to second order
perturbation theory in (\ref{gse},~\ref{dN}). This is true for
any observable that is diagonal in the basis (\ref{ons}).
However, it is
necessary to go to second order in the perturbation theory to
obtain a properly normalized ground state. 
For the eigenvalues $E_\alpha$ one finds
\begin{equation}
\label{eigenen}
E_\alpha -E_0=\left\{\matrix{
{e^2\over 2C}+\epsilon_i^{(r)}(\phi)-\epsilon_j^{(s)}\, ,\hfill
& i>N_r^+ ,\; j\leq N_s^+ \hfill\cr
{e^2\over 2C}+\epsilon_j^{(s)}-\epsilon_i^{(r)}(\phi)\, ,\hfill
& i\leq N_r^+ ,\; j>N_s^+ \hfill}\right.\, .
\end{equation}
The sums in (\ref{gse}, \ref{dN}) can be written more
explicitly in terms of the eigen-energies (\ref{eigenen}).
We obtain double sums
\begin{eqnarray}
\label{sums1}
\fl E_0-E=\sum_{i>N_r^+,\,j\leq N_s^+}
{|t_{ij}|^2 \over {e^2\over 2C}+\epsilon_i^{(r)}(\phi)-\epsilon_j^{(s)}}
+\sum_{i\leq N_r^+,\,j>N_s^+}
{|t_{ij}|^2 \over {e^2\over 2C}+\epsilon_j^{(s)}-\epsilon_i^{(r)}(\phi)} \\
\label{sums2}
\fl\Delta N_r{}^2=\sum_{i>N_r^+,\,j\leq N_s^+}
{|t_{ij}|^2 \over \left( {e^2\over 2C}
+\epsilon_i^{(r)}(\phi)-\epsilon_j^{(s)} \right)^2}
+\sum_{i\leq N_r^+,\,j>N_s^+}
{|t_{ij}|^2 \over \left( {e^2\over 2C}
+\epsilon_j^{(s)}-\epsilon_i^{(r)}(\phi) \right)^2}
\end{eqnarray}
that can be interpreted as the effect of
an electron {\em coherently} \/hopping from the stub into a virtual
state in the ring and back (first sum in the
equations~\ref{sums1},~\ref{sums2})
or vice versa (second sum). We postpone the explicit evaluation
of the above sums until after the discussion of the ensemble.

\section{\label{sec:ensemble}The ensemble}
There is little purpose
in attempting to characterize an individual sample.
If the charging energy $e^2/2C$ is small,
the critical flux $\phi^*$ is very sensitive to changes in the
particle densities $N_r/L_r$ and $N_s/L_s$.
For $e^2/2C=0$ the critical flux covers the entire interval
$[0,\phi_0/2]$ when the particle densities in the ring
and in the stub are varied independently by $\pm 1/L_r$
and $\pm 1/L_s$, respectively.
Many quantities of interest depend crucially on the number
of particles in the ring, let us mention only the persistent
current \cite{buettiker:ringstub}.
Any such quantity will depend on $\phi^*$ which itself is strongly
sample dependent. To extract more general results we
want to consider an ensemble
of rings connected to stubs and to calculate ensemble averages.    
The construction of an ensemble is thus the next task. 

We consider an ensemble of systems having different ring 
circumferences and stub lengths, but with constant
total particle number $N$, a ``strongly canonical'' ensemble
according to the classification of
Kamenev and Gefen \cite{kamenev:ensemble}.
The spectra in the ring and the stub depend on the circumference 
of the ring $L_r$
and the length of the stub $L_s$ (see equations (\ref{delta},\ref{w})). 
Therefore the topmost occupied energy levels
of the ring and the stub shift when $L_r$ and $L_s$ are varied.
We describe this shift by the difference $\Delta\epsilon$
between the energy of the topmost occupied ring state at zero flux
and the topmost occupied stub level.
For consistency with the derivation of the Coulomb energy leading
to equation (\ref{E_c}), the background charge densities $N_r/L_r$
in the ring and $N_s/L_s$ in the stub are required to be the same
for all ensemble members.  The double constraint of constant particle
number and constant densities
puts strong limits on the possible variations of $L_r$
and $L_s$.
These constraints require that the topmost
energy levels may vary at most by $w$ in energy for
the ring and by $\Delta$ for the stub.
It follows that $\Delta\epsilon$
can vary at most by $\pm(w+\Delta)$ around $0$.  Thus $\Delta\epsilon$
lies in an interval of length $2(w+\Delta)$.
We consider an ensemble with a uniform distribution of $\Delta\epsilon$
in this interval. From figure~\ref{spectrum}, one may see that at
$e^2/2C\to 0$ and if $\Delta\epsilon$
varies between $0$ and $\max\{w,\Delta\}$, $|\phi^*|$
indeed assumes any value between $0$ and $\phi_0/2$.

In the limit of vanishing interaction strength and
in the limit of a very long stub we show below 
that our system behaves effectively
like a ring coupled to a reservoir.
Hence we will refer to this limit as the ``grand canonical'' limit. 
On the other hand if the stub is very short our system 
behaves, regardless of the interaction strength, in a 
canonical manner. Hence we call the limit of a short stub also 
the canonical limit. 

In order to differentiate between quantum mechanical expectation
values and ensemble averages, we denote the latter by an overline
(e.g.\ $\overline{x}$) and the former by angular
brackets $\langle x\rangle$.

\section{Persistent current}
It is instructive to compare the Fourier
coefficients $\overline{I_n}$
of the average persistent current
\begin{equation}
\overline{I(\phi)}
=\sum_{n=1}^\infty\overline{I_n}\,\sin 2\pi n\frac{\phi}{\phi_0}
\end{equation}
with the Fourier coefficients $I_n^{(0)}$ of the
persistent current $I^{(0)}(\phi)$
of an isolated ring containing $N_r^+$ noninteracting fermions.
The $I_n^{(0)}$ are of the form
\begin{equation}
I_n^{(0)}={2e\,w\over h}{1\over n\pi}\left\{\matrix{
(-1)^n & \;\text{for $N_r^+$ odd}\hfill\cr
1 & \;\text{for $N_r^+$ even}\hfill}\right.\, .
\end{equation}
Note that the sign of the $I_n^{(0)}$ depends on
the parity of the particle number $N_r^+$
for odd $n$, whereas for even $n$ it
does not. Therefore, there is also an important difference between
the averaged Fourier coefficients $\overline{I_n}$
for odd and even $n$, respectively. We
investigate the Fourier coefficients $\overline{I_n}$
normalized with respect to the $I_n^{(0)}$:
\begin{equation}
\overline{i_n}=\frac{\overline{I_n}}{I_n^{(0)}}\, ,
\end{equation}
in particular $\overline{i_1}$ and $\overline{i_2}$.
The behaviour of these two in the limits $e^2/C\to 0$
and $e^2/C\to\infty$ is representative for
all $\overline{i_n}$ with odd and even $n$, respectively.

\begin{figure}
\centerline{\epsfysize=8cm\mbox{\epsfbox{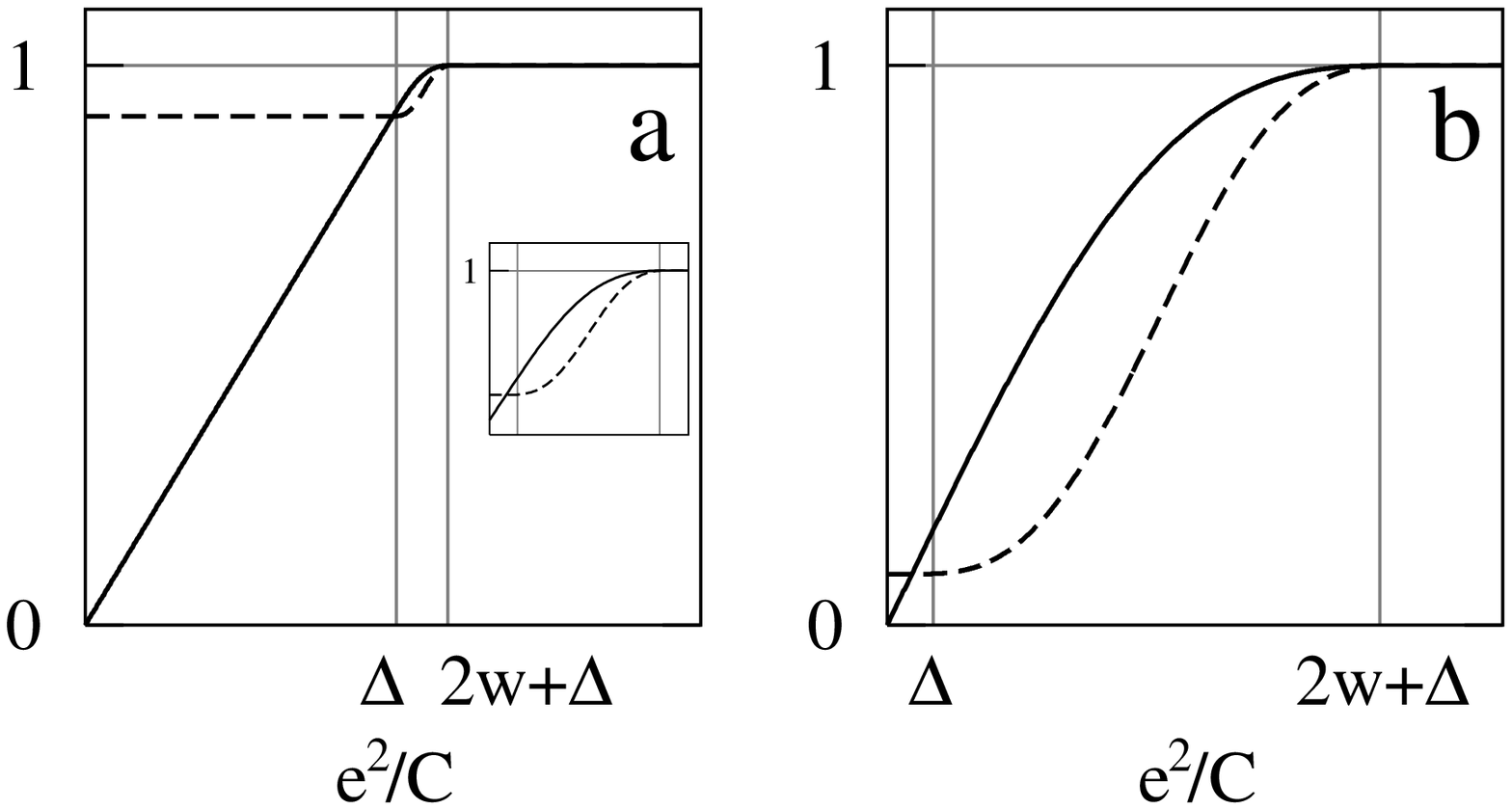}}}
\caption{\label{fcomp}Normalized Fourier coefficients of the persistent
current. The full lines stand for $\overline{i_1}$, the dashed ones
for $\overline{i_2}$. (a) shows the results for a stub that is much
shorter than the ring circumference ($L_r=10 L_s$ or equivalently
$\Delta =10w$). The inset shows the behaviour
of $\overline{i_{1,2}}$ in the crossover
region $\Delta<e^2/C<2w+\Delta$.
(b) shows the results for a stub that is much longer 
than the ring circumference: $w=10\Delta$.
In both cases $\overline{i_1}=\overline{i_2}=1$ for 
$e^2/C\geq 2w+\Delta$. For $e^2/C\leq\Delta$ one finds
that $\overline{i_1}$
goes linearly to zero whereas $\overline{i_2}=const$ in
both (a) and (b). Note that in the case (b) of the long stub,
the second Fourier coefficient is very small ($\overline{i_2}\ll 1$)
for $e^2/C\to 0$.}
\end{figure}

We obtain the $\overline{I_n}$
(and thus the $\overline{i_n}$) by calculating first the average
persistent current $\overline{I(\phi)}$
as described in section~\ref{sec:ensemble} and
then extracting them by Fourier transformation.
The first normalized Fourier
coefficient $\overline{i_1}$ vanishes linearly for
$e^2/C\to 0$ (in fact, for $e^2/C<\Delta$)
and goes to 1 for $e^2/C\geq 2w+\Delta$.
For $\overline{i_2}$ the situation is a little more complicated. It
goes to 1 as $e^2/C\geq 2w+\Delta$. For small Coulomb energies, however,
\begin{equation}
\overline{i_2}={\Delta\over w+\Delta}
\quad\text{for }\frac{e^2}{C}<\Delta\, ,
\end{equation}
i.e.\ for $w\gg\Delta$ one has $\overline{I_2}\ll I_2^{(0)}$,
whereas for $w\ll\Delta$
we find $\overline{I_2}\approx I_2^{(0)}$.
The results are sketched in figure~\ref{fcomp}.

We interpret these findings as follows. For $e^2/C$ large enough the
average persistent current becomes insensitive to the side stub. This is
true for short as well as for long stubs.
In the sense of the ``grand canonical''
and ``canonical'' limits, introduced in
section~\ref{sec:ensemble}, this indicates
that the difference between canonical and grand canonical
ensembles is less
important for interacting than for noninteracting
systems (cf.\ also~\cite{ambegaokar}).
For $e^2/C\to 0$ on the other hand, there is a difference between
short and long stubs. If the stub is long, one has $w\gg\Delta$ and
$\overline{I_2}\ll I_2^{(0)}$. As $\overline{I_1}$
vanishes in any case in this limit, the
average persistent current vanishes, too.  This
is the behaviour predicted in \cite{cheung:openring} for a ring in the
grand canonical ensemble, i.e.\ an open ring.
If $w\ll\Delta$, that is, for a short stub,
$\overline{i_2}$ does not vanish as $e^2/C$ approaches zero but
tends to a finite limit which in turn means that the persistent current
becomes $\phi_0/2$-periodic. This is what one finds for a canonical
(closed) ensemble of clean rings containing noninteracting fermions
\cite{bouchiat:canvsgrand,cheung:openring}. There is a crossover
from $\phi_0/2$-periodicity to $\phi_0$-periodicity in the energy
interval $0<e^2/C<\Delta$.

Let us point out that in the
limit $e^2/C\to 0$ the average persistent
current near $\phi =0$ is positive (it may be very small, though,
for small $\Delta$, as discussed above),
since $\overline{I_1}$ vanishes,
and $\overline{I_2}$ is always positive. Thus, in the absence of
Coulomb interactions, the ensemble shows a paramagnetic
response, a feature not observed in the experiment \cite{levy:manyrings}.
In contrast, in our system,
in the case of strong Coulomb interactions, the response of the 
ensemble can be both diamagnetic or paramagnetic. 

We have not taken into account the effect of quantum fluctuations
on the persistent current yet. This is done by taking the derivative of
(\ref{sums1}) with respect to $\phi$.
The correction is of order $|t|^2$ with respect to $\overline{I(\phi)}$.
Even for $e^2/C\geq 2w+\Delta$ it causes only
small deviations from the quasi-classical result $\overline{i_n}=1$.
The conclusions concerning the persistent current drawn from
the quasi-classical picture remain valid.
A stronger effect is seen in the particle
number fluctuations which we will discuss now.

\section{Particle number fluctuations}
\begin{figure}
\centerline{\mbox{\epsfxsize=10cm\epsfbox{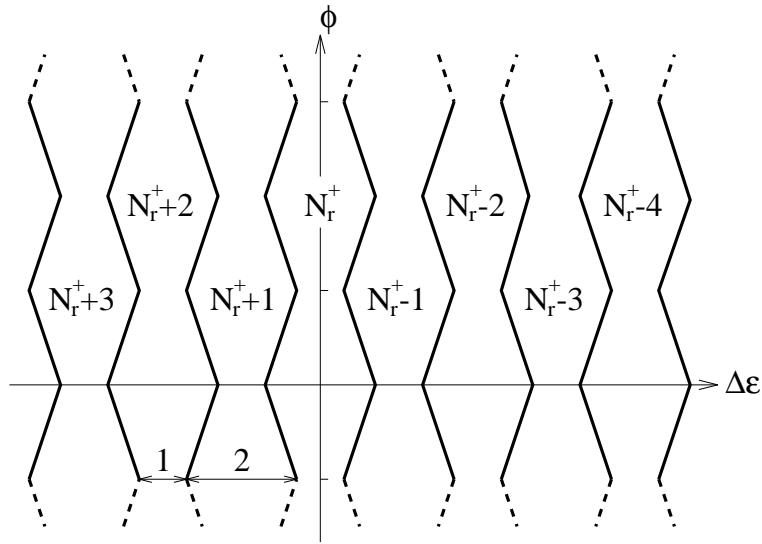}}}
\caption{\label{numbers}Domains of integer charge states in the ring
for $N_r^+$ odd, in the flux-energy plane. The boundaries of the
charge states are indicated by solid lines. The marks on the
$\phi$-axis are at $n\phi_0/2$, $n=\pm 1,\,\pm 2,\,\ldots$.
The scale on the abscissa is indicated by the lines 1 and 2, where
the length of 1 is $\Delta+e^2/C$, and
the length of 2 is $2w+\Delta+e^2/C$.}
\end{figure}

In the picture of the incoherent Coulomb blockade
(no hopping, $\hat{\Gamma}=0$),
the particle number $N_r$ in the ring can be determined
using the conditions for incoherent charge transfer
(\ref{hybridization1},~\ref{hybridization2}).
Figure~\ref{numbers} shows
the domains with definite $N_r$ in the flux-energy plane.
The particle number fluctuations can be read off this picture,
\begin{equation}
\label{deltaNqc}
\overline{(\Delta N_r{}^2)}_{inc}
\equiv\overline{N_r{}^2}-\overline{N_r}^{\,2}
=\max\left\{ 0,1-{{e^2\over C}+\Delta+2w{|\phi|\over\phi_0/2}\over 2(w+\Delta)}
\right\},
\end{equation}
the index ``inc'' referring to the incoherent Coulomb blockade
model.
This formula holds for $N_r^+$ odd; for $N_r^+$ even, the absolute 
value of the flux $|\phi|$ is replaced by 
by $\phi_0/2-|\phi|$.

\begin{figure}
\centerline{\mbox{\epsfxsize=10cm\epsfbox{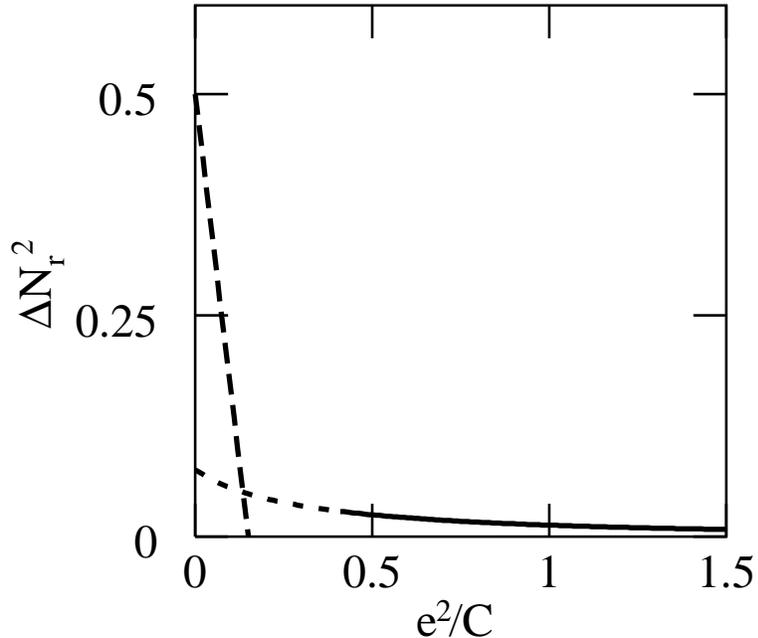}}}
\caption{\label{varN}
Particle number fluctuations as a function of $e^2/C$. The  
result in the incoherent Coulomb blockade model
$\overline{(\Delta N_r{}^2)}_{inc}$ (dashed line) and
the off-resonant result 
$\overline{\Delta N_r{}^2}$ for $e^2/C>2w+\Delta$ (solid line),
continued beyond its range of validity (short dashed line) are shown for 
$\phi=\pm\phi_0/4$, $|t|=0.02$, $w=0.1$, $\Delta=0.2$. All
energies are given in units of the band width $\Lambda$.}
\end{figure}

We shall discuss two cases, depending
on the charging energy $e^2/C$ (and on $\phi$):
\begin{enumerate}
\item\label{hop} If $e^2/C$ is sufficiently small we have
$\overline{(\Delta N_r{}^2)}_{inc}>0$. There is
real hopping and equations (\ref{hybridization1},~\ref{hybridization2})
are fulfilled for some ensemble member.
We point out that in this range of $e^2/C$ quantum contributions
to the particle number fluctuations given by equation
(\ref{deltaNqc}) are irrelevant. If we consider for a
moment an individual member of the ensemble, quantum
fluctuations are very important (of order $1$ in $|t|$) near a
resonance, see equation (\ref{numberfluct}). However, the width
in energy of the resonance is only of order $|t|$. Thus the
quantum contributions of such a resonance to the ensemble
average is only of order $|t|/(2w+\Delta)\ll 1$.
\item\label{nohop} There is only virtual hopping,
equations (\ref{hybridization1},~\ref{hybridization2})
are never fulfilled, and $\overline{(\Delta N_r{}^2)}_{inc}=0$.
We can apply the second order perturbation theory results 
from section \ref{subsec:offres}, giving (\ref{sums2}).
The sums can be explicitly evaluated e.g.\ in a symmetric narrow
band model with band width $2\Lambda$.
We assume
$\epsilon_{N_r^+}^{(r)}(\phi),\epsilon_{N_s^+}^{(s)}\gg\Lambda\gg w,\Delta$
so that we can linearize the spectra,
\begin{eqnarray}
\epsilon_{N_r^+ +k}^{(r)}(\phi)&=&\left\{\matrix{
\epsilon_{N_r^+}^{(r)}(\phi)+kw\, ,\hfill & \text{if $k$ even}\hfill\cr
\epsilon_{N_r^+ +1}^{(r)}(\phi)+(k-1)w\, ,\hfill
& \text{if $k$ odd}\hfill }\right. , \\
\epsilon_{N_s^+ +l}^{(s)}&=&\epsilon_{N_s^+}^{(s)}+l\Delta .
\end{eqnarray}
To evaluate the expression (\ref{sums2}) we shall replace the sums by
integrals and set $|t_{ij}|^2=const=|t|^2$. With these 
specifications we obtain 
\begin{equation}
\label{deltaN}
\Delta N_r{}^2\approx\left\{\matrix{
{2|t|^2\over w\Delta}\log\left({\Lambda\over e^2/2C}\right)
+\Or\left({1\over\Lambda}\right)\, ,\hfill
& \Lambda\gg{e^2\over 2C}\gg w,\Delta \hfill\cr
{2|t|^2\over w\Delta}\left({\Lambda\over e^2/2C}\right)^2
+\Or\left(\left({e^2\over 2C}\right)^{-3}\right)\, ,\hfill
& {e^2\over 2C}\gg\Lambda \hfill}\right. .
\end{equation}
Thus the result depends crucially on the relation of the 
band width $\Lambda$ to the charging energy. 
It remains to take the ensemble average, but equations (\ref{deltaN})
do not depend on the ensemble member,
thus $\overline{\Delta N_r{}^2}=\Delta N_r{}^2$.
\end{enumerate}
The behaviour of the ensemble averaged particle number fluctuations
as a function of $e^2/C$ is
summarized in figure~\ref{varN}. The incoherent fluctuations,
associated with different particle numbers
$N_r$ in the ring in different
ensemble members, and thus associated with resonant charge transfer,
vanish linearly for increasing charging energy as
discussed in (\ref{hop}). In the picture of the incoherent Coulomb
blockade a large enough charging energy therefore suppresses the
hybridization completely. A purely quantum mechanical signature
of hybridization, however, survives far into the off-resonant region,
as demonstrated by equations (\ref{deltaN}).

\section{Potential fluctuations}
The electrostatic potential on the ring is connected to the particle number
on the ring via
\begin{equation}
U_r=\frac{Q_r}{C}=\frac{e(N_r^+-N_r)}{C}.
\end{equation}
Replacing c-numbers by operators we find that it
exhibits therefore fluctuations related to the particle number
fluctuations:
\begin{equation}
\Delta U_r{}^2
\equiv\langle\hat{U_r}^2\rangle
-\langle\hat{U_r}\rangle^2
={4\over e^2}\left({e^2\over 2C}\right)^2 \Delta N_r{}^2\, .
\end{equation}
For intermediate charging energies $w,\Delta\ll e^2/2C\ll\Lambda$ we
find potential fluctuations that are {\em growing} \/with the charging
energy
\begin{equation}
\Delta U_r{}^2={8|t|^2 \over e^2\,w\Delta}
\left({e^2 \over 2C}\right)^2 \log\left({2\Lambda\over e^2/2C}\right),
\end{equation}
whereas for higher charging energies $e^2/2C\gg\Lambda$ they
tend towards a constant
\begin{equation}
\Delta U_r{}^2={8|t|^2 \Lambda^2 \over e^2\,w\Delta}\, .
\end{equation}
Within the range of validity of the perturbation theory \cite{LanLif:3},
$|t|\ll w,\Delta$, the fluctuations in the single particle spectrum
induced by potential fluctuations
$e(\Delta U_r{}^2)^{1/2}$ are in both cases much
smaller than the charging energy $e^2/2C$, but not necessarily
smaller than the level width $w$ and the level spacing in the
stub $\Delta$.

\section{Conclusions}
We present an investigation of the competing effects of
the Coulomb interaction and the hybridization
of the spectra of two coupled subsystems.
The model considered consists of a ring penetrated by a magnetic flux
coupled to a side branch. We discuss single systems as well as a
canonical ensemble. 
We identify quantities that are sensitive to
hybridization, namely the persistent current
(equations~\ref{eq:pc},~\ref{eq:pcfluct} and figure~\ref{fig:pc}) and
the particle number fluctuations in a subsystem (\ref{numberfluct})
in order to identify 
hybridization effects not only in a single- but
also in a many-particle problem.
We show that the Coulomb interaction suppresses hybridization,
abruptly in the standard incoherent Coulomb blockade model
(equation~\ref{deltaNqc} and figure~\ref{fcomp})
and smoothly when quantum effects are
taken into account (\ref{deltaN}). Interestingly, even though our 
system is strongly canonical, for small charging energies
it shows both the behaviour
of a grand canonical ensemble of rings for
a large side branch and the behaviour
of a canonical ensemble of rings for
a short side branch. Our results suggest a number of experiments 
in which either the persistent current (magnetization) or 
the charge fluctuations are measured. Such experiments can be implemented 
in mesoscopic structures or in large benzene like molecules with 
side branches.

\section*{Acknowledgments}
We thank Ya~M~Blanter and A~Levy-Yeyati for critical comments and
encouragement. This work was supported by the Swiss National Science
Foundation.

\end{document}